\documentclass[aps,prl,twocolumn,superscriptaddress]{revtex4}%
\usepackage{amsfonts}
\usepackage{amsmath}
\usepackage{amssymb}
\usepackage{graphicx}
\usepackage{xcolor}
\usepackage{epstopdf}
\usepackage{bm}%
\begin{document}
\title{Spin-Dependent High-Order Topological Insulator and Two Types of Distinct
Corner Modes in Monolayer FeSe/GdClO Heterostructure}
\author{Qing Wang}
\affiliation{Anhui Key Laboratory of Condensed Matter Physics at Extreme Conditions, High
Magnetic Field Laboratory, HFIPS, Anhui, Chinese Academy of Sciences Hefei,
230031, China}
\affiliation{Science Island Branch of Graduate School, University of Science and Technology
of China, Hefei, Anhui 230026, China}
\author{Rui Song}
\affiliation{Science and Technology on Surface Physics and Chemistry Laboratory, Mianyang,
Sichuan 621908, China}
\affiliation{Anhui Key Laboratory of Condensed Matter Physics at Extreme Conditions, High
Magnetic Field Laboratory, HFIPS, Anhui, Chinese Academy of Sciences Hefei,
230031, China}
\author{Ning Hao}
\email{haon@hmfl.ac.cn}
\affiliation{Anhui Key Laboratory of Condensed Matter Physics at Extreme Conditions, High
Magnetic Field Laboratory, HFIPS, Anhui, Chinese Academy of Sciences Hefei,
230031, China}

\begin{abstract}
We propose that a spin-dependent second-order topological insulator can be
realized in monolayer FeSe/GdClO heterostructure, in which substrate GdClO
helps to stabilize and enhance the antiferromagnetic order in FeSe. The
second-order topological insulator is free from spin-orbit coupling and
in-plane magnetic field. We also find that there exist two types of distinct
corner modes residing in intersections of two ferromagnetic edges and two
antiferromagnetic edges, respectively. The underlying physics for
ferromagnetic corner mode follows a sublattice-chirality-kink picture. More
interestingly, ferromagnetic corner mode shows spin-dependent property, which
is also robust against spin-orbit coupling. Unexpectedly, antiferromagnetic
corner mode can be taken as a typical emergent and hierarchical phenomenon
from an array of ferromagnetic corner modes. Remarkably, antiferromagnetic
corner modes violate general kink picture and can be understood as bound
states of a one-dimensional Schrodinger equation under a connected potential
well. Our findings not only provide a promising second-order topological
insulator in electronic materials, but uncover some new properties of corner
modes in high-order topological insulator.

\end{abstract}
\maketitle

In conventional $d$-dimensional topological insulator, ($d-1$)-dimensional
bulk-boundary correspondence governs the manifestation of boundary states.
Recently, this concept is generalized to ($d-n$)-dimension with $n\in
\lbrack2,d]$. The relevant quantum state is called $d$-dimensional $n$th-order
topological state, $i.e.$, high-order topological state
(HOTS)\cite{HOT-1,HOT-2,HOT-3,HOT-4,HOT-5,HOT-6,HOT-7,HOT-8,HOT-9,HOT-10,HOT-11,HOT-12}%
. As the simplest HOTS, 2D second order topological insulator (SOTI) is ideal
to study various exotic properties of HOTS, and a lot of theoretical models
are proposed for 2D SOTI. However, different from field of conventional
topological states, where topological dictionary is established and many
candidate compounds are collected, the material platforms to realize SOTI are
very limited. The current experimentally feasible platforms mainly include
photonic, phononic, acoustic and microwave- and electrical-circuit artificial
systems\cite{AH-1,AH-2,AH-3,AH-4,AH-5,AH-6,AH-7}. For the aspect of electronic
materials, some carbon-based compounds such as graphdiyne, $\gamma$-graphyne,
twisted-bilayer graphene, and bismuth heterostructure are theoretically
predicted to host 2D SOTS\cite{CA-1,CA-2,CA-3,CA-4,CA-5,CA-6,CA-7}, but have
not been experimentally realized. Furthermore, spinless or spin-polarized
feature of these proposals limits studies of spin-dependent physics of 2D
SOTI. Therefore, it remains a urgency to explore experimentally feasible
electronic materials hosting 2D SOTI, in particular, the spin-dependent 2D SOTI.

Since superconductivity with ultrahigh transition temperature ($>$65K) was
discovered in monolayer FeSe/SrTiO$_{3}$ (FeSe/STO)\cite{FeSe-1}, similar
heterostructures such as FeSe/Nb:BaTiO$_{3}$/KTaO$_{3}$, FeSe/MgO,
FeSe/AnataseTiO$_{2}$(001), and FeSe/EuTiO$_{3}$%
\cite{FeSe-2,FeSe-3,FeSe-4,FeSe-4-1} have attracted enormous interests in many
research fields. In particular, research of monolayer FeSe/STO was extended to
field of topological physics in 2014\cite{FeSe-5, FeSe-5-1}. The subsequent
theoretical work predicted a long-range Neel antiferromagnetic (AFM) order can
spontaneously form in monolayer FeSe/SrTiO$_{3}$, and a conventional
topological insulator can arise by further taking into account spin-orbit
coupling (SOC)\cite{FeSe-6}. However, whether long-range magnetic orders can
arise or not in monolayer FeSe/STO is still an experimental
debate\cite{FeSe-6,FeSe-7,FeSe-8,FeSe-9,FeSe-10,FeSe-11,FeSe-12,FeSe-13},
which reduces feasibility to realize conventional topological insulator based
on the Neel AFM in such system.

In this work, we find that a long-range Neel AFM order in monolayer FeSe can
be stabilized and enhanced through introducing a ferromagnetic (FM) insulating
substrate GdClO. Such strategy can avoid the debate of magnetism in monolayer
FeSe. The stability of the heterostructure is verified by self-consistent
first-principles calculations. Once Neel AFM order is generated with aid of
substrate GdClO, we find that a 2D SOTI naturally emerges, and is free from
SOC and in-plane magnetic field. Interestingly, the 2D SOTI hosts two types of
distinct corner modes residing in intersections of two FM edge and two AFM
edges, respectively. FM corner modes can be intrinsic or extrinsic depending
on global or local crystalline symmetry of four edge boundaries is enforced or
not. Furthermore, FM corner modes can be understood by a
sublattice-chirality-kink picture and show explicitly spin-dependent
properties. This enables one to study spin physics of 2D SOTI. Interestingly,
AFM corner modes can be taken as emergent corner modes from an array of FM
corner modes. This demonstrates an emergent and hierarchical phenomenon of
physics in a very simple and explict manner. Physically, they violate general
kink picture and correspond to the bound states of a one-dimensional
Schrodinger equation under a connected potential well. \begin{figure}[pt]
\begin{center}
\includegraphics[width=1.0\columnwidth]{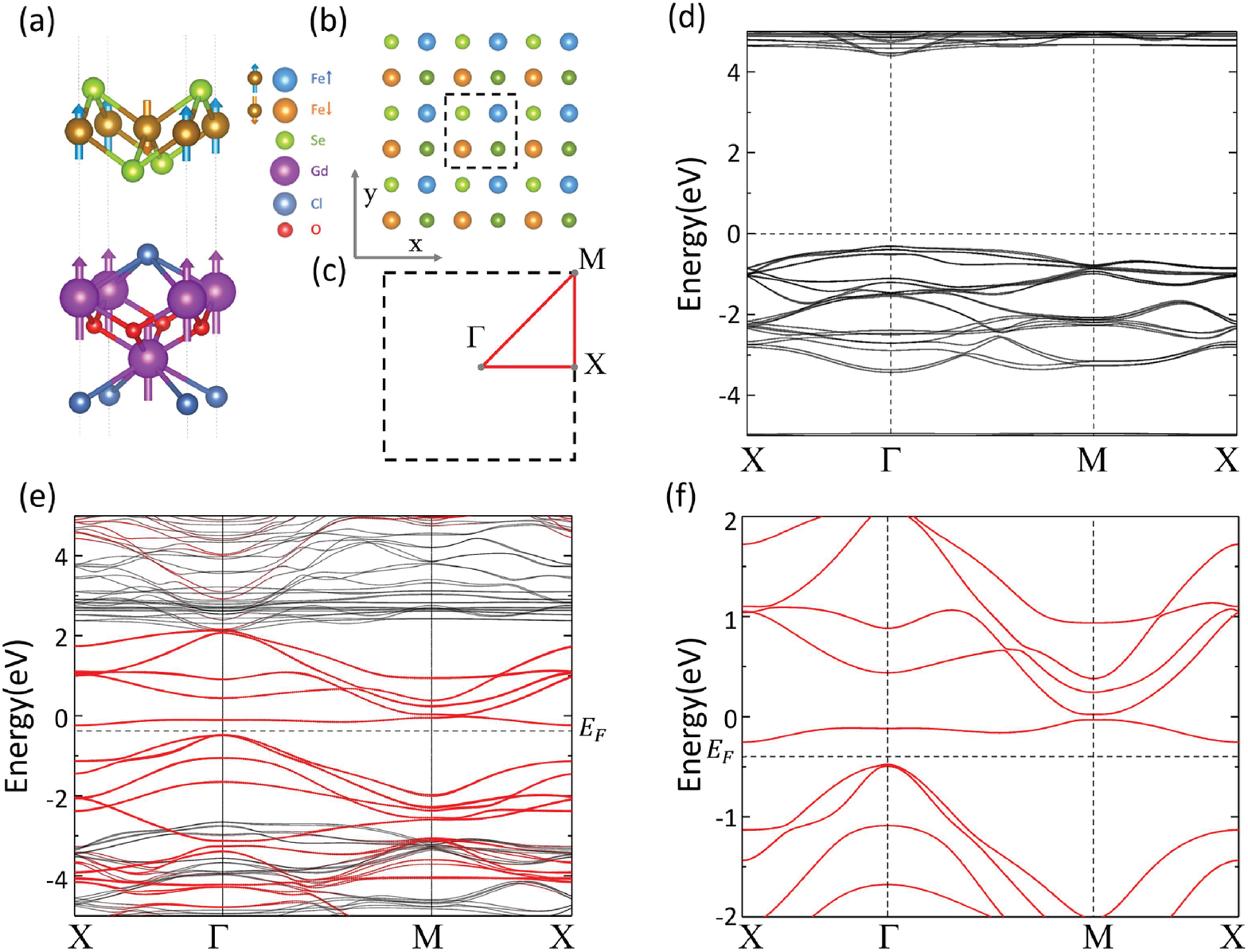}
\end{center}
\caption{(a) The spacial configuration of monolayer FeSe/GdClO
hetero-structure and the stable magnetic structure from the first-principles
calculations. (b) and (c) Lattice structure and Brillouin zone of Neel AFM
FeSe. dashed-line square in (b) labels unit cell. (d) The band structure of FM
GdClO with slab geometry. (e) The total band structure of monolayer FeSe/GdClO
heterostructure with magnetic structure in (a). (f) The bands of free-standing
monolayer FeSe by assuming a Neel AFM order. The horizontal dashed lines
labeled by $E_{F}$ in (c) and (d) denote the focused filling level.}%
\label{fig-str}%
\end{figure}

Inspired by heterostructure of monolayer FeSe/STO, the natural cleavage
surface of a promising substrate should be square lattice with matched lattice
constant to FeSe. We note that a ready compound
GdClO\cite{Gd-1,Gd-2,Gd-3,Gd-4,Gd-5} fulfills such requirements due to the
owned space group P4/nmm symmetry and matched lattice constant as same as
FeSe. Fig. \ref{fig-str} (a) shows configuration of FeSe/GdClO
heterostructure. It is well-known that GdClO is a FM insulator with a very
large gap of 5eV, as shown in Fig. \ref{fig-str} (d). The ferromagnetism is
from Gd with magnetic moment $7\mu_{B}$. Note that Gd atoms have double layer
structures, and only the top layer Gd atoms strongly couple to one sublattice
of Fe square lattices through van der Waals (vdW) interaction. It is expected
that A sublattice of Fe square lattices can arise ferromagnetism with aid of
top layer of Gd. However, the magnetism of B sublattice of Fe square lattices
cannot be intuitively determined. Thus, we perform the first-principles
calculation to consider magnetic configuration of FeSe/GdClO heterostructure.
See supplementary materials (SMs) for details. Fig. \ref{fig-str} (a) shows
the determined stable magnetic configuration. Interestingly, monolayer FeSe
arises a stable long-range Neel AFM order on a FM substrate GdClO. Fig.
\ref{fig-str} (e) gives band structures of FeSe/GdClO heterostructure. The
electronic states are nearly decoupled between monolayer FeSe and substrate
GdClO due to weak vdW interaction and non charge transfer. Fig. \ref{fig-str}
(f) gives band structure of free-standing monolayer FeSe with assumed Neel AFM
order. It is as same as FeSe part of FeSe/GdClO heterostructure shown in Fig.
\ref{fig-str} (e). Note that some recent works focused on the small gap regime
near 0eV at M point in Fig. \ref{fig-str} (f), and discussed possible 2D SOTS
induced by SOC and in-plane magnetic field\cite{FeSeAFM-1,FeSeAFM-2}. However,
the weakness of SOC and additional magnetic field fine tuning lower the
feasibility to realize the 2D SOTS and increase difficulty to study the corner
modes protected by the small SOC gap. In the following, we focus on the large
gap regime $\sim0.3$eV labeled by $E_{F}$ in Fig. \ref{fig-str} (e) and (f),
and explicitly demonstrate that a pristine 2D SOTS emerges and is free from
SOC and in-plane magnetic field.

The bulk topology of such pristine 2D SOTI can be characterized by both a
topological invariant and bulk quadrupole moment. From Fig. \ref{fig-str} (b),
monolayer Neel AFM FeSe has symmetries of $\hat{P}\hat{T}$, \{$\hat{C}%
_{2x}|1/2,0$\}, \{$\hat{C}_{4z}\hat{P}|0,1/2$\} and \{$\hat{C}_{4z}\hat
{T}|0,1/2$\} with $\hat{P}$, $\hat{T}$, \{$\hat{E}|1/2,0$\} the inversion,
time-reversal and fractional translation symmetries, respectively. The
topological invariant to characterize bulk topology of 2D SOTI can be
calculated with the method developed in a case of chiral HOTS but limited in
2D\cite{HOT-4}. Since topological invariant only depends on $\Gamma=(0,0)$ and
$M=(\pi,\pi)$ points, where the representation of fractional translation
symmetry $e^{ik_{x/y}/2}$ takes a value of $1$ or $i$ and is redundant. Then,
$(\hat{C}_{4z}\hat{P})^{4}=-1$, the eigenvalues of $\hat{C}_{4z}\hat{P}$ are
four roots of $-1$. Due to $[\hat{C}_{4z}\hat{P},\hat{P}\hat{T}]=0$ and
$\hat{P}\hat{T}$ being anti-unitary, they have to come in complex-conjugated
pairs$\{\xi e^{i\pi/4},\xi e^{-i\pi/4}\}$ with $\xi=1$ or $-1$. The
topological invariant can be defined as,%

\begin{equation}
(-1)^{v_{c}}=%
{\displaystyle\prod\limits_{n=1}^{N/2}}
\xi_{n,\Gamma}\xi_{n,M}. \label{topo_numb}%
\end{equation}
Here, $N$ labels the number of filled bands. The monolayer Neel AFM FeSe has
22 occupied bands, and calculated 11 $\xi$ values at $\Gamma/M$ points are
$\xi_{(1...11),\Gamma/M}=(\pm1,-1,\pm1,\mp1,\mp1,1,-1,1,\mp1,\mp1,\pm1)$. This
yields $v_{c}=1$ explicitly confirming a 2D SOTI. In the pioneer work of
HOTS\cite{HOT-1,HOT-2}, the high-order topology can also be understood from
change of bulk charge dipole moment $p_{x/y}$ and quadrupole moment $q_{xy}$,
which are defined as%

\begin{align}
p_{x/y}  &  =\frac{e}{2}\left(
{\displaystyle\sum\limits_{n}}
2p_{x/y}^{n}\text{ mod }2\right)  ,\label{dipole}\\
q_{xy}  &  =\frac{e}{2}\left(
{\displaystyle\sum\limits_{n}}
2p_{x}^{n}p_{y}^{n}\text{ mod }2\right)  . \label{quadrpole}%
\end{align}
Here, $p_{x/y}^{n}=q_{x/y}^{n}/2$ with $q_{x/y}^{n}$ fulfilling the equation
$(-1)^{q_{x/y}^{n}}=\eta^{n}(M)/\eta^{n}(\Gamma)$. $\eta^{n}(M/\Gamma)$
denotes the $n$th band's eigenvalue of $\hat{C}_{4z}\hat{P}$ at $M/\Gamma$
point with $\eta=$ $\xi e^{\pm i\pi/4}$. Note that $p_{x}=p_{y}$ due to
$\hat{C}_{4z}\hat{P}$ symmetry, and the summation is over all the occupied
bands. The 22 (11 pairs) eigenvalues of $\hat{C}_{4z}\hat{P}$ symmetry at
$M/\Gamma$ point are listed in SMs. Then, $(p_{x},p_{y})=0$ and $q_{xy}=e/2$,
respectively. This supports a 2D SOTI in monolayer Neel AFM FeSe/GdClO.
\begin{figure}[pt]
\begin{center}
\includegraphics[width=1.0\columnwidth]{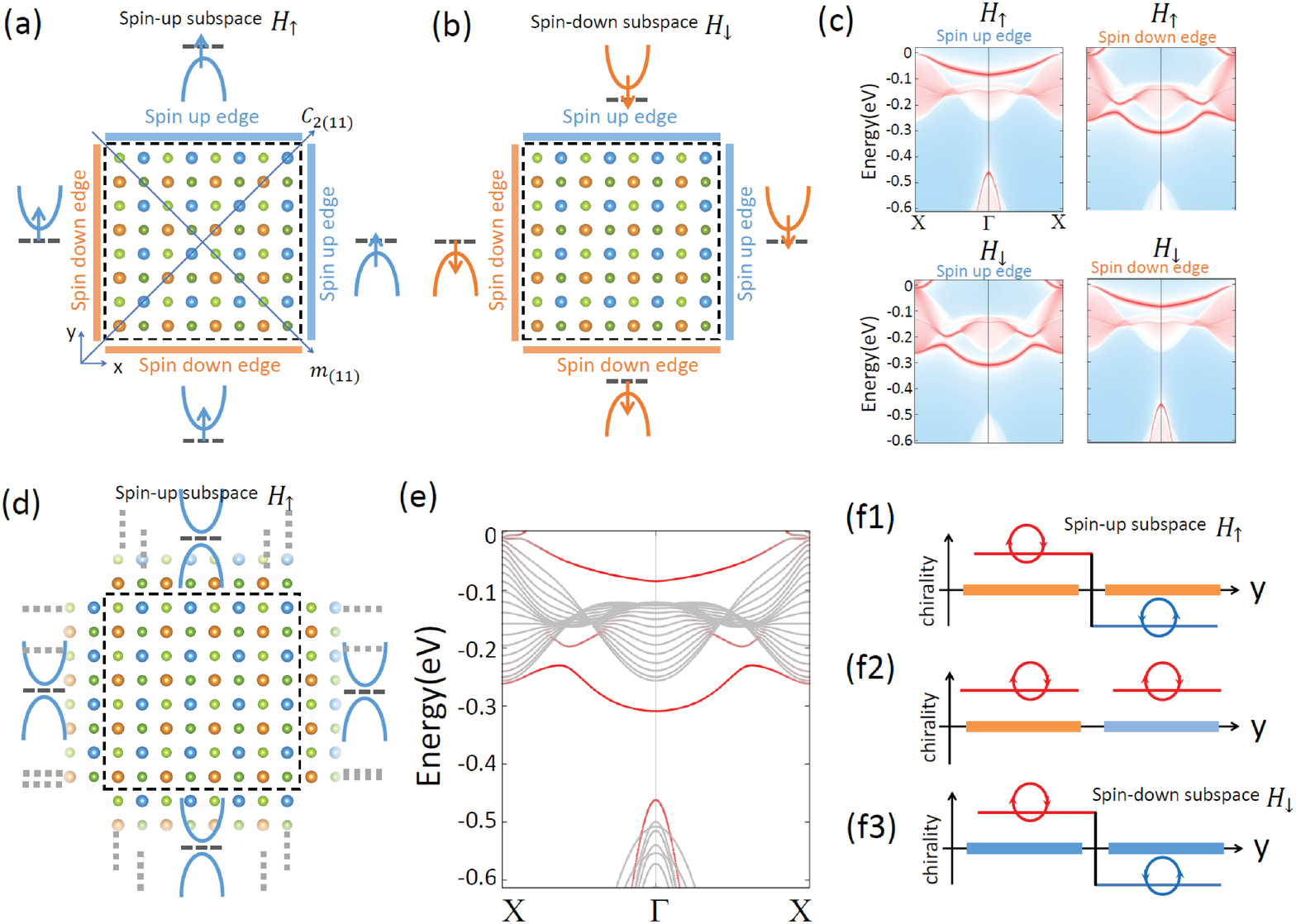}
\end{center}
\caption{(a)-(b) Illustration of monolayer Neel AFM FeSe cluster with size of
4$\times$4 in unit of unit cells shown in Fig. 1(b). (c) The edge spectrum of
spin-up and -down Hamiltonian $H_{\uparrow}$ and $H_{\downarrow}$,
respectively. The relevant spin-polarized half massive Dirac bands are
schematically plotted near four edges of cluster in (a) and (b). (d) The full
massive Dirac bands can be obtained by sticking two parallel edges
corresponding to (a). (e) The full edge spectrum without spin-resolved
property. (f1)-(f3) Three kinds of sublattice chirality kinks for three types
of corners in (a) and (b). }%
\label{fig-FM}%
\end{figure}\ 

\begin{figure}[pt]
\begin{center}
\includegraphics[width=1.0\columnwidth]{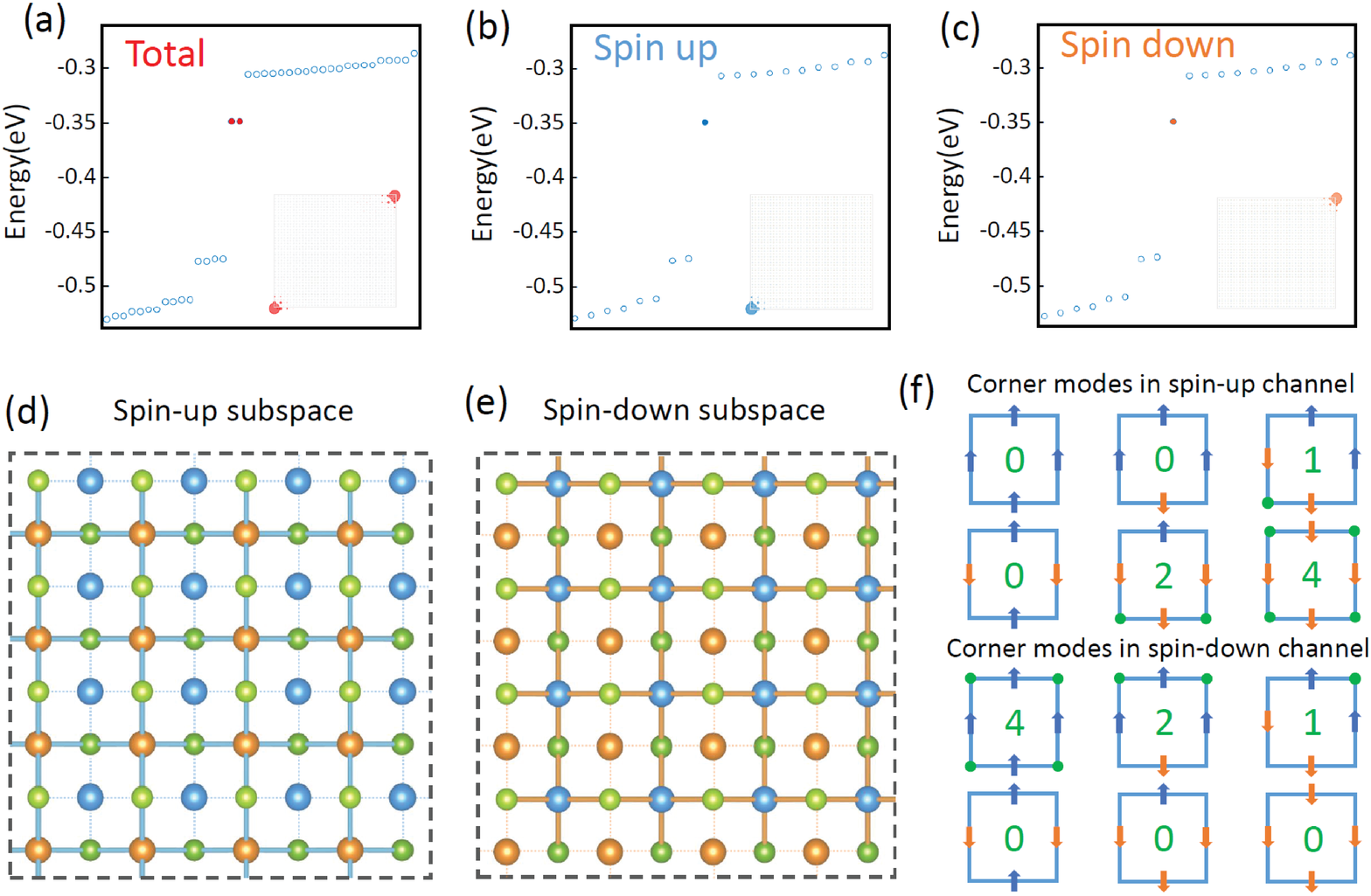}
\end{center}
\caption{(a)-(c) The discrete energy spectrum of cluster in Fig. 2 (a) without
SOC. (a) for no spin-resolved case and (b), (c) for spin-dependent cases. The
inserts in (a)-(c) give the density distribution of the corner modes. The
cluster size takes 20$\times$20 in calculations. (d) and (e) The relative
hopping integral patterns between Fe xz/yz/z$^{2}$ and Se z orbital for two
spin-decoupled Hamiltonian without SOC, respectively. The strong and weak
hoppings are labeled by thick and thin connections, respectively. (f) The
summary of distribution of corner modes with definite spin-up and -down
polarization in several different of clusters, respectively. }%
\label{fig-spin}%
\end{figure}

From the generalized bulk-boundary correspondence, a 2D SOTI enables existence
of 0d corner modes. Fig. \ref{fig-FM} (a) shows a typical cluster of monolayer
Neel AFM FeSe. It has two kinds of corners formed by two edges with the same
and opposite FM orders, respectively. The existence of corner modes can be
understood under following sublattice-chirality-kink picture. From edge
spectrum in Fig. \ref{fig-FM} (c) and (e), two spatially seperated one
dimensional massive Dirac bands at two parallel $x$ or $y$ edges can be
recombined by sticking two paralled edges togrther in each spin polarization
subspace, as shown in Fig. \ref{fig-FM} (a) and (d). Then, a low-energy
effective Hamiltonian to describe the edge spectrum can be expressed as%

\begin{align}
H_{\uparrow,x/y}(k_{x/y}) &  =\pm vk_{x/y}\tau_{y/x}+m\tau_{z},\label{low-x}\\
H_{\downarrow,x/y}(k_{x/y}) &  =\pm vk_{x/y}\tau_{y/x}-m\tau_{z},\label{low-y}%
\end{align}
Here, $v$ is an effective velocity. $k_{x/y}$ are momenta along $x/y$
direction, $\tau_{x/y/z}$ are three Pauli matrices defined in sublattice space
with intertwined orbital degree of freedom. $m$ is the mass induced by the AFM
order. The cluster in Fig. \ref{fig-FM} (a) has $\hat{P}\hat{T}$, $\hat
{C}_{2(1,1)}\hat{T}$ and $\hat{m}_{(1,1)}$ symmetries, and two different
corners formed by intersections of two same and opposite FM edges,
respectively. Note that $\hat{C}_{2(1,1)}\hat{T}$ builds intra-spin-subspace
connection, and $\hat{P}\hat{T}$ and $\hat{m}_{(1,1)}$ constructs
inter-spin-subspace connection. Consider orbital weight of bulk bands near
$E_{F}$ in Fig. \ref{fig-str} (e), the basis functions of Hamiltonian in Eqs.
(\ref{low-x}) and (\ref{low-y}) can be defined as $\psi_{\uparrow
}(k)=[d_{A,z^{2},\uparrow}(k),d_{B,xz+iyz,\uparrow}(k)]^{T}$ \cite{FeSe-5,
FeSe-5-1} and $\psi_{\downarrow}=\hat{P}\hat{T}\psi_{\uparrow}(k)$ (See Fig.
S1 in SMs). Then, the representation matrix $U$ of $\hat{C}_{2(1,1)}\hat{T}$
is $\frac{1}{2}[(1+\tau_{z})-i(1-\tau_{z})]\mathcal{K}$ with $\mathcal{K}$ the
complex conjugate. For instance, in spin-up subspace, $UH_{\uparrow,x}%
(k_{x})U^{\dag}|_{k_{x}\rightarrow-k_{y}}=$ $vk_{y}\tau_{x}+m\tau_{z}$. In
comparison with $H_{\uparrow,y}(k_{y})$, the sublattice chirality defined by
$\vec{k}\times\vec{\tau}$ changes sign. Namely, there exists a
sublattice-chirality kink, as shown in Fig. \ref{fig-FM} (f1). Thus, a corner
mode must appear for the corner formed by two same FM edges \cite{Zero-mode}.
However, for the corner formed by two opposite FM edges, $H_{\uparrow,x}%
(k_{x})\overset{\hat{m}_{(1,1)}}{\longrightarrow}$ $H_{\downarrow,y}%
(k_{y})\overset{\hat{P}\hat{T}}{\longrightarrow}$ $H_{\uparrow,y}(k_{y})$.
Namely, two edge Hamiltonian are identical under $\hat{m}_{(1,1)}$ in the
restricted $H_{\uparrow}(k)$ subspace, as shown in Fig. \ref{fig-FM} (f2).
Thus, no zero mode will emerge. Similarly, in spin-down subspace, another
corner mode appears from the sublattice-chirality-kink picture, as shown in
Fig. \ref{fig-FM} (f3). Fig. \ref{fig-spin} (a) gives the spectrum of the
cluster. One can find two degenerate corner modes located on two corners from
two same FM edges. Clusters with other patterns and the relevant spectrum are
shown in SMs. Note that corner modes are robust against SOC and in-plane
magnetic field (See Fig. S3 and S4 in SMs). Different from robustness of
topological boundary states in conventional topological insulator, the corner
modes in 2D SOTI depend on details of patterns of the clusters (See Fig. S2 in
SMs). Thus, the role of crystalline symmetry is subtle. The 2D SOTI can be
intrinsic if four edges is globally considered and $\hat{C}_{2(1,1)}\hat{T}$
is enforced. Otherwise, the 2D SOTI can be extrinsic if two edges are
considered locally\cite{HOT-6,Edge-1,Edge-2}. In any case, two FM edges
connected by $\hat{C}_{2(1,1)}\hat{T}$ is the key to enable appearance of
corner mode.

Different from previous proposals, above spin-dependent
sublattice-chirality-kink picture indicates the FM corner modes should also be
spin-dependent. Fig. \ref{fig-spin} (b) and (c) give the spin-dependent
spectrum and distribution of the spin-polarized corner modes. The location of
corner modes can be understood from Fig. \ref{fig-spin} (d) and (e), where the
strong and weak hopping integrals are schematically plotted\ in spin-up and
-down subspace, respectively (See Fig. S6\ in SMs). The strong bonds of iron
atoms at left-lower and top right corners are broken in spin-up and -down
subspace, respectively. This leaves the relevant corners with isolated
spin-dependent corner modes in Fig. \ref{fig-spin} (b)-(e). Note that the
fidelity of spin polarization is 99.6\% with SOC, because spin-dependent
nature is governed by magnetic splitting with energy scale 2.5eV in comparison
to tiny SOC energy $\sim$ 0.03eV. It indicates spin-dependent feature is
robust against other weak external perturbations and enables ones to
manipulate spin degree of freedom of corner modes in possible application. In
Fig. \ref{fig-spin} (f), we list various possibilities to realize corner modes
in two spin channels.

\begin{figure}[pt]
\begin{center}
\includegraphics[width=1.0\columnwidth]{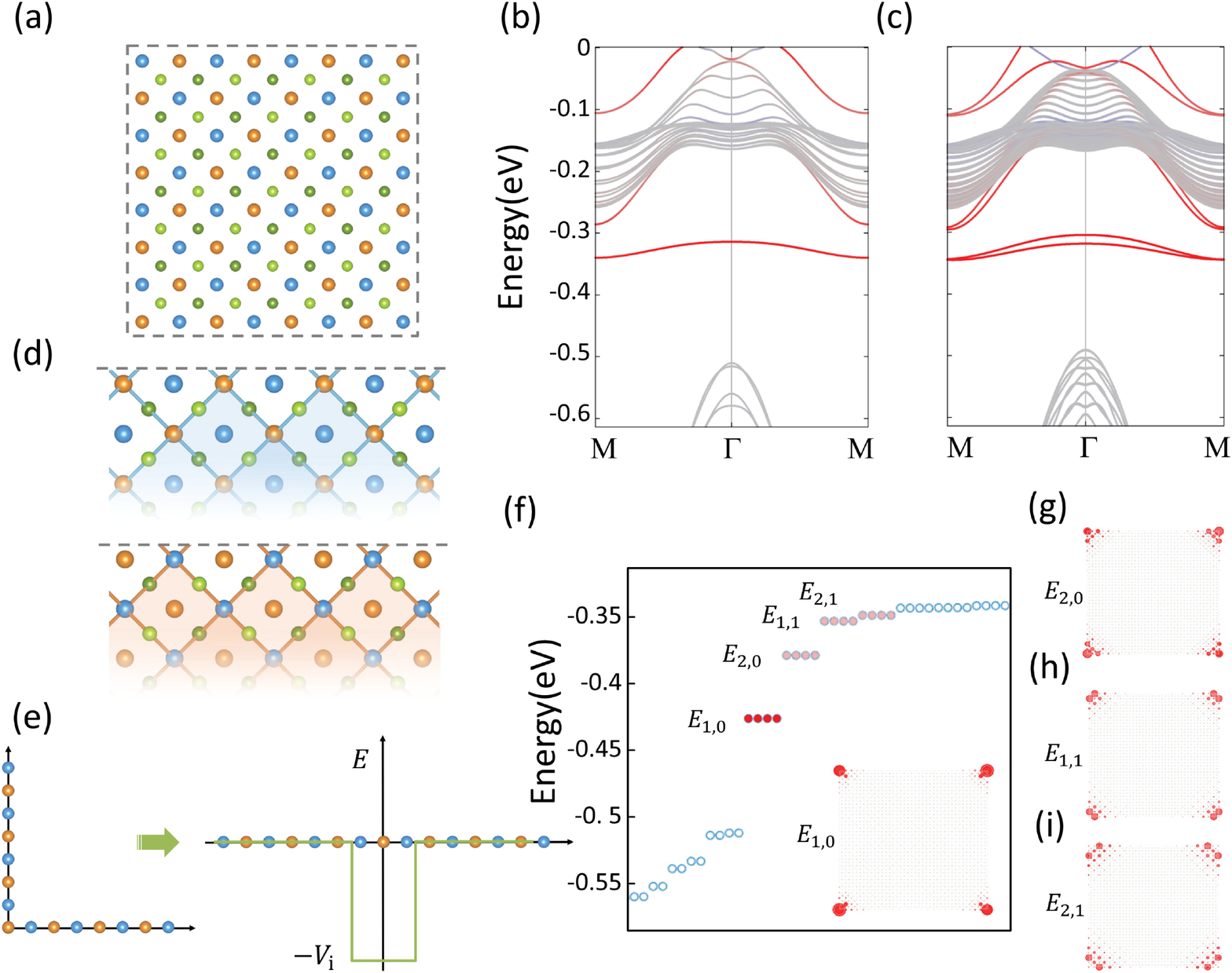}
\end{center}
\caption{(a) A cluster of monolayer FeSe with size 4$\times$4 in unit of
$\sqrt{2}$ unit cell shown in Fig. 1(b). The cluster size is 14$\times$14 in
calculation of (e)-(h). (b)-(c) The AFM edge spectrum without SOC and with
SOC, respectively. (d)The relative hopping integral patterns for decoupled
spin-up (top panel) and -down (bottom panel) Hamiltonian without SOC,
respectively. (e) The mapping of two AFM edges from orthogonal to linear
configuration . (f) The discrete energy spectrum corresponds to (a). The
insert gives the density distribution of the corner mode with four-quadruple
degeneracy and the lowest energy. (f)-(h) The density distribution of other
three corner modes with increasing energy. }%
\label{fig-AFM}%
\end{figure}

The cluster of FeSe/GdClO can have distinct AFM edges as shown in Fig.
\ref{fig-AFM} (a). Fig. \ref{fig-AFM} (b) and (c) give the relevant edge
spectrum. The remarkable feature is emergence of two nearly flat bands near
$E_{F}\sim-0.35$eV. Their formation can be understood according to Fig.
\ref{fig-AFM} (d), in which the strong and weak hopping integrals are labeled
in spin-up and -down subspace, respectively. Then, each spin-down and -up iron
atom with broken strong bonds can bound a corner mode, as shown in top and
bottom panels in Fig. \ref{fig-AFM} (d), respectively. The isolated corner
modes arrange to one-dimensional array and has weak coupling to form two
degenerate flat bands shown in Fig. \ref{fig-AFM} (b). SOC can further induce
weak coupling to break the degeneracy of flat bands shown in Fig.
\ref{fig-AFM} (c). Numerical calculations for square cluster with four AFM
edges indicate there exist four different AFM corner modes with each four
quadruple degeneracies, as shown in Fig. \ref{fig-AFM} (f). The AFM corner
modes have lower energy than bottom of flat bands. It indicates the general
kink picture is violated. Interestingly, we find that AFM corner modes can be
understood by following one-dimensional Schrodinger equation through
considering the edge mapping shown in Fig. \ref{fig-AFM} (e),
\begin{equation}
H_{AFM}(k\rightarrow-i\partial_{x})\psi(x)=E\psi(x),\label{EqAFM-1}%
\end{equation}
with%
\begin{equation}
H_{AFM}(k\rightarrow-i\partial_{x})=-\frac{\hbar^{2}}{2m^{\ast}}\frac{d^{2}%
}{dx^{2}}+V(x).\label{EqAFM-2}%
\end{equation}
Here, $m^{\ast}$ is effective mass of flat bands. $V(x)$ is
effective potential well near a corner of two AFM edges. $V(x)$ can be
approximately expressed by a square potential well. $V(x)=-V_{i}$ when
$\left\vert x\right\vert <a/2$ and otherwise $V(x)=0$. Here, $a$ denotes width
of the potential. $i$ is 1 or 2 to label two different flat bands. The
bound-state solutions of Eq. (\ref{EqAFM-1}) are standard. The even-parity
solution is $E_{i,n}=\frac{\pi^{2}\hbar^{2}}{2m^{\ast}a^{2}}(2n+1)^{2}-V_{i}$
and $\psi_{n}(x)\sim\cos k_{n}x$ for $\left\vert x\right\vert <a/2$ and $\sim
e^{-\alpha_{n}\left\vert x\right\vert }$ for $\left\vert x\right\vert >a/2$.
The odd-parity solution is $E_{i,n}=\frac{\pi^{2}\hbar^{2}}{2m^{\ast}a^{2}%
}(2n+2)^{2}-V_{i}$ and $\psi_{n}(x)\sim\sin k_{n}x$ for $\left\vert
x\right\vert <a/2$ and $\sim e^{-\alpha_{n}\left\vert x\right\vert }$ for
$\left\vert x\right\vert >a/2$. $n$ takes 0, 1, 2... The numerical results in
Fig. \ref{fig-AFM} (f) indicate the emergent four different corner modes
correspond to $E_{1,0}$, $E_{2,0}$, $E_{1,1}$, $E_{2,1}$ with consistent even
and odd alternating distribution patterns as shown in Fig. \ref{fig-AFM}
(f)-(i). It seems that the AFM corner modes are not topologically protected.
However, their emergence can be stretched back to FM corner modes. Thus, both
FM and AFM corner modes are topologically protected in a different
hierarchical manner. Note that there exists at least a AFM corner mode for any
$V(x)$ well.

In many previous proposals, in-plane external magnetic field or magnetic
proximity effect is induced to drive 2D SOTI. The corner modes are sensitive
to these fine tunings. It improves difficulties to construct electronic
devices based on corner modes. The 2D SOTI and relevant corner modes here are
robust against in-plane external magnetic field, and other perturbations (See
Fig. S3-S5 in SMs). The corner modes here lie at 0.35eV below original Fermi
energy. To experimentally detect them, scanning tunnelling spectroscopy is a
feasible technique, which can measure spin-resolved local density of states by
changing voltage in a large regime. The transport properties of the corner
modes can be studied only when their energy is tuned to the Fermi level. Such
tuning can be experimentally realized by electrostatic gating or proton gating
techniques, both of which are mature and have especial advantages in layered
heterostructures\cite{Gate-1,Gate-2,Gate-3}. In particular, such gated tuning
has been realized in FeSe thin flakes\cite{Gate-4,Gate-5,Gate-6}. Once
appropriate hole carriers are induced by gating to move corner mode energy to
Fermi energy, monolayer FeSe/GdClO could be an ideal platform to study the
properties of 2D SOTI and relevant corner modes.

In conclusion, we propose that a monolayer FeSe/GdClO heterostructure can
realize a 2D second-order topological insulator, which is free from SOC and
in-plane magnetic field, Furthermore, we find that corner modes are protected
by a large gap of about 0.3eV, and can be detected at high temperature. We
also find there exist two distinct types of FM and AFM corner modes. More
interestingly, we show that FM corner modes follow a sublattice-chirality-kink
picture and have unique spin-dependent property, and AFM corner modes emerge
from FM corner mode array. The diversity of FM and AFM corner modes provide
new way to construct the relevant devices by utilizing their spin degree of freedom.

\begin{acknowledgments}
We thank Lin Hao for helpful discussions. This work was financially supported
by National Natural Science Foundation of China under Grants (No. 12022413,
No. 11674331), the National Key R\&D Program of China (Grant No.
2017YFA0303201), the \textquotedblleft Strategic Priority Research Program
(B)\textquotedblright\ of the Chinese Academy of Sciences, Grant No.
XDB33030100, the Collaborative Innovation Program of Hefei Science Center, CAS
(Grants No. 2020HSC-CIP002), the Major Basic Program of Natural Science
Foundation of Shandong Province (Grant No. ZR2021ZD01). A portion of this work
was supported by the High Magnetic Field Laboratory of Anhui Province, China.
\end{acknowledgments}


\begin{thebibliography}{99}                                                                                               %


\bibitem {HOT-1}W. A. Benalcazar, B. A. Bernevig, and T. L. Hughes, Electric
multipole moments, topological multipole moment pumping, and chiral hinge
states in crystalline insulators, Phys. Rev. B \textbf{96}, 245115 (2017).

\bibitem {HOT-2}W. A. Benalcazar, B. A. Bernevig, and T. L. Hughes, Quantized
electric multipole insulators, Science \textbf{357}, 61-66 (2017).

\bibitem {HOT-3}Z. Song, Z. Fang, and C. Fang, (d-2)-dimensional edge states
of rotation symmetry protected topological states, Phys. Rev. Lett.
\textbf{119}, 246402 (2017).

\bibitem {HOT-4}F. Schindler, A. M. Cook, M. G. Vergniory, Z. Wang, S. S. P.
Parkin, B. A. Bernevig, and T. Neupert, Higher-order topological insulators,
Science Advances 4, 10.1126/sciadv.aat0346 (2018).

\bibitem {HOT-5}G. van Miert and C. Ortix, Higher-order topological insulators
protected by inversion and rotoinversion symmetries, Phys. Rev. B \textbf{98},
081110 (2018).

\bibitem {HOT-6}E. Khalaf, Higher-order topological insulators and
superconductors protected by inversion symmetry, Phys. Rev. B \textbf{97},
205136 (2018).

\bibitem {HOT-7}M. Geier, L. Trifunovic, M. Hoskam, and P. W. Brouwer,
Second-order topological insulators and superconductors with an order-two
crystalline symmetry, Phys. Rev. B \textbf{97}, 205135 (2018).

\bibitem {HOT-8}D. C\u{a}lug\u{a}ru, V. Juri\v{c}i\'{c}, and B. Roy,
Higher-order topological phases: A general principle of construction, Phys.
Rev. B \textbf{99}, 041301 (2019).

\bibitem {HOT-9}Y. Ren, Z. Qiao, and Q. Niu, Engineering corner states from
two-dimensional topological insulators, Phys. Rev. Lett. \textbf{124}, 166804 (2020).

\bibitem {HOT-10}C.-A. Li and S.-S. Wu, Topological states in generalized
electric quadrupole insulators, Phys. Rev. B \textbf{101}, 195309 (2020).

\bibitem {HOT-11}Y.-B. Yang, K. Li, L.-M. Duan, and Y. Xu, Type-ii quadrupole
topological insulators, Phys. Rev. Research \textbf{2}, 033029 (2020).

\bibitem {HOT-12}M. Jung, Y. Yu, and G. Shvets, Exact higher-order
bulk-boundary correspondence of corner-localized states, Phys. Rev. B
\textbf{104}, 195437 (2021).

\bibitem {AH-1}S. Mittal, V. V. Orre, G. Zhu, M. A. Gorlach, A. Poddubny, and
M. Hafezi, Photonic quadrupole topological phases, Nature Photonics
\textbf{13}, 692 (2019).

\bibitem {AH-2}L. He, Z. Addison, E. J. Mele, and B. Zhen, Quadrupole
topological photonic crystals, Nature Communications \textbf{11}, 3119 (2020).

\bibitem {AH-3}M. Serra-Garcia, V. Peri, R. Susstrunk, O. R. Bilal, T. Larsen,
L. G. Villanueva, and S. D. Huber, Observation of a phononic quadrupole
topological insulator, Nature 555, 342 (2018).

\bibitem {AH-4}Y. Qi, C. Qiu, M. Xiao, H. He, M. Ke, and Z. Liu, Acoustic
realization of quadrupole topological insulators, Phys. Rev. Lett.
\textbf{124}, 206601 (2020).

\bibitem {AH-5}C. W. Peterson, W. A. Benalcazar, T. L. Hughes, and G. Bahl, A
quantized microwave quadrupole insulator with topologically protected corner
states, Nature \textbf{555}, 346 (2018).

\bibitem {AH-6}S. Imhof, C. Berger, F. Bayer, J. Brehm, L. W. Molenkamp, T.
Kiessling, F. Schindler, C. H. Lee, M. Greiter, T. Neupert, and R. Thomale,
Topolectricalcircuit realization of topological corner modes, Nature Physics
\textbf{14}, 925 (2018).

\bibitem {AH-7}M. Serra-Garcia, R. Susstrunk, and S. D. Huber, Observation of
quadrupole transitions and edge mode topology in an lc circuit network, Phys.
Rev. B 99, 020304 (2019).

\bibitem {CA-1}E. Lee, R. Kim and J. Ahn et al. Two-dimensional higher-order
topology in monolayer graphdiyne. npj Quantum Materials \textbf{5}, 1-7 (2020).

\bibitem {CA-2}X. Sheng, C. Chen and H. Liu et al. Two-dimensional
second-order topological insulator in graphdiyne. Phys. Rev. Lett.
\textbf{123}, 256402 (2019).

\bibitem {CA-3}B. Liu, G. Zhao and Z. Liu et al. Two-dimensional quadrupole
topological insulator in $\gamma$-graphyne. Nano. Lett. \textbf{19}, 6492-6497 (2019).

\bibitem {CA-4}M. Park, Y. Kim and G. Cho et al. Higher-order topological
insulator in twisted bilayer graphene. Phys. Rev. Lett. \textbf{123}, 216803 (2019).

\bibitem {CA-5}C. Ma, Q. Wang and S. Mills et al. Moir\'{e} band topology in
twisted bilayer graphene. Nano. Lett. \textbf{20}, 6076-6083 (2020).

\bibitem {CA-6}B. Liu, L. Xian and H. Mu et al. Higher-order band topology in
twisted moir\'{e} superlattice. Phys. Rev. Lett. \textbf{126}, 066401 (2021).

\bibitem {CA-7}C. Chen, Z. Song, J. Zhao, Z. Chen, Z. Yu, X. Sheng and S. A.
Yang Universal Approach to Magnetic Second-Order Topological Insulator, Phys.
Rev. Lett. \textbf{125}, 056402 (2020).

\bibitem {FeSe-1}Q. Y. Wang, Z. Li, W. H. Zhang, Z. C. Zhang, J. S. Zhang, W.
Li, H. Ding, Y. B. Ou, P. Deng, K. Chang et al., Interface Induced High
Temperature Superconductivity in Single Unit-Cell FeSe Films on SrTiO$_{3}$,
Chin. Phys. Lett. \textbf{29}, 037402 (2012).

\bibitem {FeSe-2}R. Peng, H. C. Xu, S. Y. Tan, H. Y. Cao, M. Xia, X. P. Shen,
Z. C. Huang, C.H.P. Wen, Q. Song, T. Zhang, B. P. Xie, X. G. Gong and D. L.
Feng, Tuning the band structure and superconductivity in single-layer FeSe by
interface engineering, Nat. Commun. \textbf{5}, 5044 (2014).

\bibitem {FeSe-3}J. Shiogai, Y. Ito, T. Mitsuhashi, T. Nojima and A.
Tsukazaki, Electric-field-induced superconductivity in electrochemically
etched ultrathin FeSe films on SrTiO$_{3}$ and MgO, Nat. Phys. \textbf{12},
42--46 (2016).

\bibitem {FeSe-4}H. Ding, Y. Lv, K. Zhao, W. Wang, L. Wang, C. Song, X. Chen,
X. Ma, and Q. Xue, High-Temperature Superconductivity in Single-Unit-Cell FeSe
Films on Anatase TiO2(001), Phys. Rev. Lett. \textbf{117}, 067001 (2016).

\bibitem {FeSe-4-1}C. Liu, H. Shin, A. Doll, H. Kung, Ryan P. Day, B. A.
Davidson, J. Dreiser, G. Levy, A. Damascelli, C. Piamonteze, and K. Zou,
High-temperature superconductivity and its robustness against magnetic
polarization in monolayer FeSe on EuTiO3, npj Quantum Mater. \textbf{6} 85 (2021).

\bibitem {FeSe-5}N. Hao and J. Hu, Topological Phases in the Single-Layer
FeSe, Phys. Rev. X \textbf{4}, 031053 (2014).

\bibitem {FeSe-5-1}N. Hao and J. Hu, Topological quantum states of matter in
iron-based superconductors: from concept to material realization, Natl. Sci.
Rev. 6, 213-226 (2019).

\bibitem {FeSe-6}Z. F.Wang,, H. Zhang, D. Liu, C. Liu, C. Tang, C. Song, Y.
Zhong, J. Peng, F. Li, C. Nie, L. Wang, X. Zhou, X. Ma., Q. Xue and F. Liu,
Nat. Mater. \textbf{15}, 968 (2016).

\bibitem {FeSe-7}F. Zheng, Z. Wang, W. Kang and P. Zhang, Antiferromagnetic
FeSe monolayer on SrTiO$_{3}$: The charge doping and electric field effects,
Sci Rep \textbf{3}, 2213 (2013).

\bibitem {FeSe-8}K. Liu, Z. Lu, and T. Xiang, Atomic and electronic structures
of FeSe monolayer and bilayer thin films on SrTiO$_{3}$ (001):
First-principles study, Phys. Rev. B \textbf{85}, 235123 (2012)

\bibitem {FeSe-9}H. Cao, S. Chen, H. Xiang, and X. Gong, Antiferromagnetic
ground state with pair-checkerboard order in FeSe, Phys. Rev. B 91, 020504(R) (2015).

\bibitem {FeSe-10}S. Y. Tan, et al. Interface-induced superconductivity and
strain-dependent spin density wave in FeSe/SrTiO$_{3}$ thin films. Nat. Mater.
\textbf{12}, 634--640 (2013).

\bibitem {FeSe-11}H. Y.Cao, S. Y. Tan, H. J. Xiang, D. L. Feng, and X. G.
Gong, Interfacial effects on the spin density wave in FeSe/SrTiO3 thin films.
Phys. Rev. B \textbf{89}, 014501 (2014).

\bibitem {FeSe-12}Y. Zhou, L. Miao, P. Wang, F.\thinspace F. Zhu, W.\thinspace
X. Jiang, S.\thinspace W. Jiang, Y. Zhang, B. Lei, X.\thinspace H. Chen,
H.\thinspace F. Ding, Hao Zheng, W.\thinspace T. Zhang, Jin-feng Jia, Dong
Qian, and D. Wu, Antiferromagnetic Order in Epitaxial FeSe Films on

SrTiO$_{3}$, Phys. Rev. Lett. \textbf{120}, 097001 (2018).

\bibitem {FeSe-13}Y. Gu, Q. Wang, H. Wo, Z. He, H. C. Walker, J. T. Park, M.
Enderle, A. D. Christianson, W. Wang, J. Zhao, Frustrated magnetic
interactions in FeSe, arXiv:2207.10981 (2022).

\bibitem {Gd-1}C. Bungenstock, T. Tr\"{o}ster, W. B. Holzapfel, L. Fini and M.
Santoro, Energy levels of Pr$^{3+}$:GdOCl under pressure, J. Phys.: Condens.
Matter \textbf{12} 6959, (2000).

\bibitem {Gd-2}K. R. Kort and S. Banerjee, Oriented Electrophoretic Deposition
of GdOCl Nanoplatelets, J. Phys. Chem. B \textbf{117}, 1585-1591 (2013).

\bibitem {Gd-3}C. Bungenstock, T. Tr\"{o}ster, W. B. Holzapfel, Effect of
pressure on free-ion and crystal-field parameters of Pr$^{3+}$in LOCl (L=La,
Pr, Gd), Phys. Rev. B \textbf{62}, 7945 (2000).

\bibitem {Gd-4}M. R. Osanloo1, M. L. Van de Put, A. Saadat and W. G.
Vandenberghe, Identification of two-dimensional layered dielectrics from first
principles, Nat. Commun. \textbf{12}, 5051 (2021).

\bibitem {Gd-5}D. Boglaienkoa, A. Andersenb, S. M. Healdc, T. Vargab, D. R.
Mortensend, S. Tetefe, G. T. Seidlere, N. Govindf, T. G. Levitskaiaa, X-ray
absorption spectroscopy of trivalent Eu, Gd, Tb, and Dy chlorides and
oxychlorides, Journal of Alloys and Compounds, \textbf{897}, 162629 (2022).

\bibitem {FeSeAFM-1}Aiyun Luo, Zhida Song and Gang Xu, Fragile topological
band in the checkerboard antiferromagnetic monolayer FeSe, npj Comput. Mater.
\textbf{8}, 26 (2022)

\bibitem {FeSeAFM-2}Haimen Mu, Gan Zhao, Huimin Zhang and Zhengfei Wang,
Antiferromagnetic second-order topological insulator with fractional
mass-kink, npj Comput. Mater. \textbf{8}, 82 (2022).

\bibitem {Zero-mode}R. Jackiw and C. Rebbi, Solitons with fermion number 1/2,
Phys. Rev. D 13, 3398 (1976).

\bibitem {Edge-1}E. Khalaf, W. A. Benalcazar, T. L. Hughes, and R. Queiroz,
Boundary-obstructed topological phases, Phys. Rev. Research \textbf{3}, 013239 (2021).

\bibitem {Edge-2}L. Trifunovic and P. W. Brouwer, Higher-Order Topological
Band Structures, Phys. Status Solidi B \textbf{258}, 2000090 (2021).

\bibitem {Gate-1}J. T. Ye, S. Inoue, K. Kobayashi, Y. Kasahara, H. T. Yuan, H.
Shimotani and Y. Iwasa, Liquid-gated interface superconductivity on an
atomically flat film, Nat. Mater. \textbf{9}, 125--128 (2009).

\bibitem {Gate-2}J. T. Ye , Y. J. Zhang, R. Akashi, M. S. Bahramy, R. Arita,
and Y. Iwasa, Superconducting dome in a gate-tuned band insulator, Science
\textbf{338}, 1193--1196 (2012).

\bibitem {Gate-3}D. Costanzo, S. Jo, H. Berger and A. F. Morpurgo,
Gate-induced superconductivity in atomically thin MoS2 crystals, Nat.
Nanotechnol. \textbf{11}, 339--344 (2016).

\bibitem {Gate-4}B. Lei, J.\thinspace H. Cui, Z.\thinspace J. Xiang, C. Shang,
N.\thinspace Z. Wang, G.\thinspace J. Ye, X.\thinspace G. Luo, T. Wu, Z. Sun,
and X.\thinspace H. Chen, Evolution of high-temperature superconductivity from
a low-Tc phase tuned by carrier concentration in FeSe thin flakes. Phys. Rev.
Lett. 116, 077002 (2016).

\bibitem {Gate-5}B. Lei, N. Z. Wang, C. Shang, F. B. Meng, L. K. Ma, X. G.
Luo, T. Wu, Z. Sun, Y. Wang, Z. Jiang, B. H. Mao, Z. Liu, Y. J. Yu, Y. B.
Zhang, and X. H. Chen, Tuning phase transitions in FeSe thin flakes by
field-effect transistor with solid ion conductor as the gate dielectric. Phys.
Rev. B \textbf{95}, 020503 (2017).

\bibitem {Gate-6}C. S. Zhu, J. H. Cui, B. Lei, N. Z. Wang, C. Shang, F. B.
Meng, L. K. Ma, X. G. Luo, T. Wu, Z. Sun, and X. H. Chen, Tuning electronic
properties of FeSe$_{0.5}$Te$_{0.5}$ thin flakes using a solid ion conductor
field-effect transistor. Phys. Rev. B \textbf{95}, 174513 (2017).
\end{thebibliography}
\end{document}